
\magnification 1200
\baselineskip=6 truemm
\vsize=22.5 truecm
\voffset 0.5 truecm
\hoffset 1 truecm
\hsize=14.5 truecm
\nopagenumbers
\def\ref{\par\noindent\hangindent 1.0 truecm}
\null\vskip 1.0truecm

\centerline {\bf PRELIMINARY RESULTS FROM THE}
\centerline {\bf ESO SLICE PROJECT (ESP)}
\bigskip
G. Vettolani and E. Zucca
\item\item{\it Istituto di Radioastronomia del CNR, Bologna, Italy}
\par
A. Cappi, R. Merighi, M. Mignoli, G. Stirpe and G. Zamorani
\item\item{\it Osservatorio Astronomico di Bologna, Bologna, Italy}
\par
C. Collins and H. MacGillivray
\item\item{\it Royal Observatory Edinburgh, Edinburgh, United Kingdom}
\par
C. Balkowski, J. Alimi, A. Blanchard, V. Cayatte,
\par
P. Felenbok, S. Maurogordato and D. Proust
\item\item{\it DAEC, Observatoire de Paris--Meudon, Meudon, France}
\par
G. Chincarini and L. Guzzo
\item\item{\it Osservatorio Astronomico di Brera, Milano, Italy}
\par
D. Maccagni
\item\item{\it Istituto di Fisica Cosmica e Tecnologie Relative, Milano, Italy}
\par
R. Scaramella
\item\item{\it Osservatorio Astronomico di Roma, Monteporzio Catone, Italy}
\par
M. Ramella
\item\item{\it Osservatorio Astronomico di Trieste, Trieste, Italy}
\bigskip\noindent
{\bf Abstract}
\par
We present the first results of a galaxy redshift survey,
ESO Slice Project (ESP),
we are accomplishing as an ESO Key--Project
over about 40 square degrees in a region near the
South Galactic Pole. The limiting magnitude is $b_J = 19.4$.
Up to now $\sim 85\%$ of the observations has been
completed and  $\sim 65\%$ of the data has been reduced
providing $\sim 2000$ galaxy redshifts.
\par
We present some preliminary results concerning the large scale galaxy
distribution, the luminosity function and the properties of galaxies with
spectra showing prominent emission lines.

\medskip\noindent
{\bf 1. Introduction}
\medskip
\par
In September 1991 we started a galaxy redshift survey over a strip of
$22^{\circ}
\times 1^{\circ}$ (plus a near--by area of $5^{\circ} \times 1^{\circ}$,
five degrees
west of the strip) in the South Galactic Pole region.
The right ascension limits are $ 22^{h} 30^m$ and
 $ 01^{h} 20^m $, at a mean declination of $ -40^{\circ}$ (1950).
We have filled this area with a regular grid of circular fields
with a diameter of $32\ arcminutes$ (see Figure 1). This size corresponds to
the field of view of the multifiber spectrograph (OPTOPUS) we use at the
ESO $3.6m$ telescope.
\par
The limiting magnitude of the survey is $b_J \leq 19.4$. We have selected
the target objects from the Edinburgh--Duhram Southern Sky Galaxy
Catalogue (Heydon--Dumbleton et al. 1989), which has been obtained from
COSMOS scans of 60 plates in the SGP region. The catalogue has a $95 \%$
completeness at $b_J \leq 20.5$, a photometric accuracy
of $\sim 0.03$ magnitudes and an estimated stellar contamination
$\leq 10 \%$.
\par
The total number of objects we have  observed is of the order of 4000.
We have obtained spectra for galaxies in all the fields and have
reduced the data for 70 of them (shown as filled
circles in Figure 1), obtaining 1924 galaxy redshifts.
\par
At $z \simeq 0.1$, which corresponds to the peak of the selection function
of the survey, the linear dimensions of the full strip are of the order of
$110 \times 5\  h^{-1}\ Mpc$.
\topinsert\nobreak
\vskip 4truecm
\nobreak
\par
\item\item{{\bf Fig.1:}~~The survey plan: each circle corresponds to
one OPTOPUS field. The spectra of galaxies inside the fields shown as
filled circles have been already  reduced.}
\bigskip\endinsert
The main goals of our project can be summarized as follows:
\par\noindent
{\bf 1)} Determination of the  galaxy luminosity function
(estimating both its shape and normalization) in a volume with dimensions
large enough to average over the large scale inhomogeneities.
\par\noindent
{\bf 2)} Study of the statistics of emission line galaxies (both normal and
active) in a large, unbiased sample of galaxies.
\par\noindent
{\bf 3)} Measure of the size distribution of inhomogeneities in the
galaxy distribution over a large volume.
\medskip\noindent
{\bf 2. Observations and Data Reduction}
\medskip
\par
The observations have been obtained with the multifiber spectrograph OPTOPUS
at the Cassegrain focus of the ESO $3.6m$ telescope at La Silla. OPTOPUS has 50
fibers which are manually plugged into holes drilled in aluminum plates.
The diameter on the sky of the fibers is $2.4 ~arcseconds$.
At $b_{J} \leq 19.4$ the median number of galaxies per OPTOPUS
field is 34 and $85\%$ of the fields has less than 45 galaxies. Five
fibers are dedicated to the sky observations and the remaining 45 fibers
are used to observe galaxies. When the number of galaxies in the
field exceeds the number of available fibers, the galaxies to be
observed are chosen at random from the galaxy list. However, the
overdense fields have been often observed twice to ensure
a completeness of the order of $80 \%$ or larger.
\par
Under optimal conditions we have observed up to 6 or 7 fields per night.
The exposure time for each field is one hour,
split into two 30 minute exposures in order to ease the cosmic rays removal.
Comparison spectra and white lamp flats have been taken at the beginning
and  end of each exposure.
The spectra cover the wavelength range from 3730 \AA~~to 6050 \AA, with an
effective resolution of about 3.5 \AA.
\par
Spectra are flat--fielded, extracted and wavelength calibrated. Then the
relative transmission of each fiber
is computed by normalizing each spectrum through the flux of the
OI $\lambda 5577$ and Na $\lambda 5891$ sky lines.
\par
After subtraction of the average sky from the galaxy plus sky spectra
(the accuracy of the sky subtraction is in the range 2--$5\%$),
the redshifts are measured by cross--correlating
the spectra with a set of template stars observed by us with the same
instrument. Redshift measurements obtained from the cross--correlation
are not considered reliable when the value of the $R$ parameter
(Tonry \& Davis 1979) is  $ R \le 3$, unless the measurement is confirmed
by emission lines at the same redshift.
The redshifts from emission lines (when present) are also measured, as well as
the equivalent width of the most common emission lines.
The median error in velocity for the total sample is $\sim 50\ km/s$.
\par

\topinsert\nobreak
\vskip 9truecm
\nobreak
\par
\item\item{{\bf Fig.2:}~~Wedge diagram in the region
$22^h 30^m <\alpha< 1^h 30^m$ and $-41^o 15' <\delta<-39^o 45'$. }
\bigskip\endinsert

Figure 2 shows the wedge diagram of the 1924 galaxies with measured redshift.
About $80\%$ of the fields observed up to now have a redshift
completeness greater than $60\%$. The less complete fields have been
reobserved, but the new data have not been reduced yet.
\medskip\noindent
{\bf 3. Large Scale Properties}
\medskip
\par
The histogram in Figure 3a shows the distribution in comoving distance
($q_{\circ} = 0.5$) of the
1924 galaxies measured so far.
The outstanding peak at $ D \simeq 300\ h^{-1}\ Mpc$
is not due to a single galaxy cluster but to a structure which extends over
almost all the fields (see Figure 2).
 The distance of this peak is almost coincident
with the maximum of our selection function. Other peaks seen in Figure
3a at smaller and larger distance  are due to less prominent structures,
but with a similar contrast with
respect to a uniform distribution. These structures are clearly seen
in Figure 4, which shows
a better representation of the large scale distribution of galaxies in
this survey. This figure has been produced by giving
to  each galaxy in each field a weight proportional to
the selection function (derived from the
luminosity function described in the next
paragraph) and  to the incompleteness of the field it
belongs to. This procedure implicitly assumes
 that the galaxies we did not observe
follow the same distribution
in depth as the observed galaxies. This assumption is justified by the
fact that, in overdense fields, the galaxies we observed were selected
at random and that the apparent magnitude distributions of the galaxies with
and without redshift are fully compatible.
Data were subsequently smoothed with a gaussian window of $6\ h^{-1}\ Mpc$.

\topinsert\nobreak
\vskip 12.0truecm
\nobreak
\par
\item\item{{\bf Fig.3:}~~a)~~Galaxy distribution in comoving distance
($q_0 = 0.5$).
The solid line shows the expectation resulting from a uniform distribution
of the galaxies in the sample. Vertical lines show the position of
the BEKS peaks (see text).}
\item\item{b)~~Galaxy distribution in comoving distance;
the hatched histogram displays only galaxies showing emission lines.}
\bigskip\endinsert

\topinsert\nobreak
\vskip 11.0truecm
\nobreak
\par
\item\item{{\bf Fig.4:}~~Wedge diagram showing the  galaxy distribution
in the first row of OPTOPUS fields,
smoothed with a gaussian window function, taking into
account both the incompleteness and the selection function of
the sample.  The $X$ and
$Y$ scales are in $h^{-1}\ Mpc$ (comoving). Note that East and West are
flipped with respect
to the previous wedge diagram.}
\bigskip\endinsert

\par
A surprisingly large fraction of galaxies ($\sim 40\%$) shows the
presence of one or more emission lines (OII $\lambda 3727$, $H\beta$,
OIII $\lambda 4959$ and $\lambda 5007$).
These objects can be either spiral galaxies, where lines originate
mostly from HII regions in the disks, or early--type galaxies undergoing
a significant burst of star formation.
\par
The distribution in distance of the galaxies with emission lines is
shown as a dashed histogram in Figure 3b.
 It is clear from the figure and also confirmed by statistical tests,
that the large scale distribution
of galaxies with emission lines
is different from that of galaxies without
emission lines:  the observed peaks in the galaxy
distribution are much less
apparent in galaxies with emission lines.
This suggests that either spiral galaxies are less frequent in the
densest regions, thus confirming a large scale validity of the well known
morphology--density relation, or starbursts phenomena occur preferentially
in low density environments or both.
\par
After completion of the survey we will be able to explore further this point
on the basis of quantitative estimates of densities and a clearer
definition of the
structures beyond pure visual impression.
\par
The vertical lines in Figure 3a show the location of the regularly
spaced density enhancements as Broadhurst et al. (1990) found in
their deep pencil beam survey in the South Galactic Pole region.
The two main peaks in our redshift distribution (at comoving distances
of $\sim 170$ and $\sim 300\ h^{-1}\ Mpc$) are in reasonably good agreement
with the Broadhurst et al. (1990) peaks and may well be part of the same
structures (walls) orthogonal to the line of sight.
\par
Under this hypothesis, since the Broadhurst et al. (1988)
survey region is located
$\sim 10^{\circ}$ north of the eastern corner of the present survey,
the structure at $z \simeq  0.1 $ would have minimum linear dimensions
of the order of $ 110 \times 50\ h^{-1}\ Mpc$, comparable with the Great Wall
(Geller \& Huchra 1989).
\medskip\noindent
{\bf 4. The Luminosity Function}
\medskip
\par
The well controlled selection of our sample and the large number
of redshifts already obtained allow us to estimate for the first time the
shape parameters of the galaxy luminosity function at magnitudes as faint as
$b_J = 19.4$.
\par
Since our database was selected in the blue--green, K--corrections are needed
to compute the luminosity function even for the moderate redshifts sampled
by our galaxies ($z \leq 0.2$).
The functional forms of the K--correction as a function of redshift have
been taken from Shanks et al.
(1984). Application of the K--correction to each individual galaxy
obviously requires the knowledge of its morphological type.
In our case the determination of the morphological types
by visual inspection of the plates is difficult since most of our
galaxies are fainter than
$17^{th}$ magnitude. To overcome this problem, we have adopted the following
statistical approach.
First, we have assumed that the percentages of late-- and early--type galaxies
in our sample are approximately the same as those observed in brighter and
nearer samples, {\it i.e.} $\sim 70 \%$ and $\sim 30\%$ respectively
(Shanks et al. 1984). Then, we have applied the K--correction appropriate
for a late--type galaxy to all galaxies showing emission lines ($\sim 40 \%$) .
Finally, we have randomly assigned
a morphological type to the remaining  galaxies ($\sim 60\%$)
in such a way as to obtain the assumed ratio of late and early types in
the total sample. A number of tests performed by varying this ratio
show that the final results are only mildly dependent on the adopted
value.
\par
Following the STY method of Sandage et al. (1979), we  have then derived
the parameters of
the Schechter functional form of the luminosity function,
through a maximum likelihood technique.
The best fit parameters are $\alpha = -1.13 $ and $M^*_{b_J} = -19.70 $.
\par
In order to estimate the maximum amount of uncertainty induced by our
``statistical''  K--correction scheme, we have also computed the two
parameters of the luminosity function with the extreme assumptions that
all galaxies are either ellipticals or spirals. The best fit parameters
obtained
in these way are $\alpha =-1.15 $, $M^*_{b_J} = -19.87 $ and
$\alpha = -1.05 $, $M^*_{b_J} = -19.56 $, respectively.

\topinsert\nobreak
\vskip 11.0truecm
\nobreak
\item\item{{\bf Fig.5:}~~Galaxy luminosity function: solid circles
have been computed
using a modified version of the C--method and the normalization is arbitrary. }
\bigskip\endinsert

\par
Figure 5 shows the luminosity function (with arbitrary normalization)
obtained for our sample:
the solid line corresponds to the Schechter functional form obtained
through the STY method, while the
solid circles have been determined through
a modified version of the non--parametric C--Method (Lynden--Bell 1971).
The 68\% and 95\% confidence ellipses for
$\alpha$ and $M^*_{b_J}$ are shown in the inset, where the cross represents
the best estimate obtained by Loveday et al. (1992) from their survey of
galaxies with $b_J \leq 17.15$.

\topinsert\nobreak
\vskip 11.0truecm
\nobreak
\item\item{{\bf Fig.6:}~~68\% and 95\% confidence ellipses for
$\alpha$ and $M^*_{b_J}$ for galaxies with and without detectable
emission lines.  }
\bigskip\endinsert

\par
The excellent agreement between our best fit parameters and those obtained
by Loveday et al. shows that, up to a depth of $\simeq 600\ h^{-1}\ Mpc$,
the overall shape of the luminosity function is well
described by a Schechter function with a faint--end slope $\alpha \sim
-1.1$ up to $M = -16$.
\par
Figure 6 shows the 68\% and 95\% confidence ellipses for
$\alpha$ and $M^*_{b_J}$ for galaxies with and without detectable
emission lines. The allowed contours are clearly separated at a very high
significance level: the galaxies with emission lines have steeper
power law slopes and fainter $M^*_{b_J}$. This result is much stronger and
qualitatively consistent with the differences in the parameters of the
luminosity functions of early-- and late--type galaxies found
by Loveday et al. (1992).
\medskip\noindent
{\bf 5. Star Forming Galaxies}
\medskip
\par
 The OII doublet, $ \lambda 3727$, is the most useful star
formation tracer in the blue since this line can serve,
although with larger uncertainty,
as a substitute of $H\alpha$ (Kennicut 1992).
 If $R$ is the number of newly formed stars in solar masses per year,
$EW$ the equivalent width of the
 OII line (in \AA) , and $L$ the galaxy blue luminosity in solar units,  then
$$ R = 7 \times 10^{-12} L \times EW \times h^{-2} $$
  (Kennicut 1992).
 \par
The measurement of the equivalent width of the detected emission lines is
still incomplete; the following results
refer to $\sim 1/3 $ of the presently available sample.
Moreover a statistical analysis of the emission line properties
 is further complicated by the fact that the  line detectability varies as
 a function of the spectrum signal to noise ratio. To overcome this problem
 we have proceeded as follows.
 \par
  Simple considerations on the line detectability allows to correlate
  the signal to noise in the line $(S/N)_l$ to its observed equivalent
  width $ EW $ and the signal to noise in the adiacent continuum
  $ (S/N)_c $ as:
  $$  (S/N)_l \simeq 0.4 \times EW \times (S/N)_c $$
The   signal to noise ratio for the continuum has been evaluated for all the
objects showing emission lines in two regions of $ 100$ \AA~~each at the blue
and red sides of the OII line, and the resulting data are shown in Figure 7.
The relation corresponding to  $(S/N)_l = 10$ is also shown.
This relation  clearly defines an envelope below which very few lines have been
identified
visually and measured. We therefore assume that $(S/N)_l \sim 10$ is
our detection limit.
  For galaxies not showing emission lines, $(S/N)_c$ has been
  evaluated at the expected position of the OII line on the basis of the
absorption
lines redshift. This  allows
 to estimate the upper limit of the OII equivalent width
through the above relation.

\topinsert\nobreak
\vskip 12.0truecm
\nobreak
\par
\item\item{{\bf Fig.7:}~~Distribution of the observed OII equivalent widths
versus signal
to noise ratio of the adiacent continuum. The solid line represents a signal to
noise in the line $ (S/N)_l = 10$.}
\bigskip\endinsert

\par
 We have, therefore, equivalent widths (or corresponding upper
limits) of OII $\lambda 3727$ for the whole sample, allowing detailed
 statistical analysis from $z = 0$ to $ z \simeq 0.3$.
\par
The first relevant question we asked is : is there any systematic
dependence of OII equivalent width (which is proportional to
the number of new stars per unit luminosity)  with galaxy luminosity and/or
 redshift
(which is roughly proportional to the lookback time at low $z$)?
\par
Application of a bivariate regression analysis (ASURV package,
Isobe \& Feigelson 1986) to our data (both detections and upper limits)
shows that both these correlations are present but with different levels of
significance. While the correlation of $EW$ with absolute
magnitude ($EW$ decreasing with increasing luminosity) is significant
at  $ 4.5 \sigma$, the correlation with $z$ ($EW$ increasing with increasing
$z$)
is significant at  $2.5 \sigma$ only.
\par
Similar results are obtained by studying the cumulative distribution of $EW$ as
a function of $M$ in fixed $z$ bins or as a function of $z$
in fixed $M$ bins.
For instance, we have studied the cumulative distributions of OII $EW$
 for galaxies of different luminosities
in two bins in redshift $( 0.13 \leq z \leq 0.185$ and $ z \leq 0.09 )$.
With this choice of the redshift bins we are not considering for this analysis
the galaxies located in the main peak of the density distribution, where
the fraction of emission lines is significatively lower than the average.
In both redshift bins
faint galaxies have larger OII equivalent widths than bright ones.
Part of this effect might be induced by a different mixture of
morphological types as a function of absolute magnitude,
with the fraction of spirals increasing
toward the faint end of the luminosity function; however, the fact that this
correlation is significant also for
galaxies brighter than $ M_{b_J} = -18.8$, where the
ratio of spirals to ellipticals is approximately constant,
makes us believe that the correlation is true also within a given morphological
type.
\par
The dependence of $EW$ versus redshift
 is more difficult to study for two reasons. First, the redshift interval
covered by our data is not very large; second, as already mentioned,
the  redshift distribution  shows several peaks,
where a lack of emission line galaxies is found (see Figure 3b).
Preliminary results seem to suggest that for a fixed magnitude interval
the correlation between OII $EW$ and redshift is not very significant.
This is in agreement with the results of the bivariate regression analysis
mentioned above.
These correlations of OII equivalent width with redshift and/or luminosity
will be discussed in greater detail when the whole data set will be available.
\bigskip
\noindent
 {\bf  Acknowledgements}
\par
This work,  based on data collected at the European Southern Observatory,
has been partially supported  through NATO Grant CRG 920150 and EEC Contract
ERB--CHRX--CT92--0033.
\bigskip
\noindent
{\bf References}
\medskip
\ref
Broadhurst, T.J., Ellis, R.S., Shanks, T., 1988,  MNRAS 235 827
\ref
Broadhurst, T.J., Ellis, R.S., Koo, D.C., Szalay, A.S., 1990,
Nature 343 726
\ref
Geller, M.J., Huchra, J.P., 1989, Science 246, 897
\ref
Isobe,T., Feigelson, E.D., 1986, Ap.J. 31, 209
\ref
Heydon--Dumbleton, N.H., Collins, C.A., MacGillivray, H.T., 1988, in
{ \it Large--Scale Structures in the Universe}, ed. W. Seitter, H.W. Duerbeck
and M. Tacke, Springer--Verlag, p. 71
\ref
Kennicut, R.C., 1992, Ap.J. 388, 310
\ref
Loveday, J., Peterson, B.A., Efstathiou, G., Maddox, S.J., 1992,
Ap.J. 390 338
\ref
Lynden--Bell, D., 1971, MNRAS 155 95
\ref
Sandage, A., Tamman, G., Yahil, A., 1979, Ap.J. 232 252
\ref
Shanks, T., Stevenson, P.R.F., Fong, R., MacGillivray, H.T., 1984,
MNRAS 206 767
\ref
Tonry, J., Davis, M., 1979, Ap.J. 84 1511
\vfill
\hfill
\eject
\bye